\RequirePackage{fix-cm}
\documentclass{amsart}                     
\usepackage[foot]{amsaddr}
\usepackage[table,xcdraw]{xcolor}
\usepackage{graphicx}
\usepackage{amsmath}
\usepackage{amssymb}
\usepackage{adjustbox}
\usepackage{multirow}
\usepackage{tikz}
\usepackage{xfrac}
\usepackage[hang,small,bf]{caption}    
\usetikzlibrary{backgrounds,fit,decorations.pathreplacing}  
\newcommand{\ket}[1]{\ensuremath{\left|#1\right\rangle}} 

\usepackage{pgfplots}

\usepackage[utf8]{inputenc}
\usepackage{tikz}   
\usetikzlibrary{arrows}
\usepackage[most]{tcolorbox}
\usepackage{filecontents}
\usepackage{pgfplots}

\usepackage{tkz-euclide}
\usetkzobj{all}

\usepackage{subfig}
\usepackage{booktabs}
\usepackage{pst-node}

\usepackage{latexsym}
\begin{document}
\pagenumbering{arabic}
\title{Quantum one-way permutation over the finite field of two elements}
\author{Alexandre de Castro¹}

\address{¹Laborat\'{o}rio de Matem\'{a}tica Computacional, Centro Nacional de Pesquisa Tecnol\'{o}gica em Inform\'{a}tica para a 
              Agricultura (Embrapa Inform\'{a}tica Agropecu\'{a}ria), Empresa Brasileira de Pesquisa Agropecu\'{a}ria, 13083-886 Campinas-SP, Brazil}
\email{alexandre.castro@embrapa.br}         

\maketitle
\begin{abstract}
In quantum cryptography, a one-way permutation is a bounded unitary operator $U:\mathcal{H} \to \mathcal{H}$ on a Hilbert space $\mathcal{H}$ that is easy to
compute on every input, but hard to invert given the image of a random input. Levin [Probl. Inf. Transm., vol. 39 (1): 92-103 (2003)] has conjectured that the unitary transformation 
$g(a,x)=(a,f(x)+ax)$, where $f$ is any length-preserving function and $a,x \in GF_{{2}^{\|x\|}}$, is an information-theoretically secure operator 
within a polynomial factor. Here, we show that Levin's one-way permutation is provably secure because its output values
are four maximally entangled two-qubit states, and whose probability of factoring them 
approaches zero faster than the multiplicative inverse of any positive polynomial $poly(x)$ over the Boolean ring of all subsets of $x$. Our results demonstrate through 
well-known theorems that existence of classical one-way functions implies existence of a universal quantum one-way permutation that cannot be inverted in subexponential time in the worst case.
\end{abstract}
\section{Introduction.}
One of the remarkable effects of (pseudo)randomness is breaking the symmetries inherent in many natural and artificial phenomena \cite{Levin2005}. 
Because one-way permutations are quite heavily involved in the generation of (pseudo)randomness, they are seen as (pseudo)random generators themselves 
\cite{GoldreichLevin1989}. In the following, we will analyze Levin's construction \cite{Levin2003} that addresses the existence of a specific one-way 
permutation, a one-to-one and onto mapping whose probability of security failure is negligible for a
cryptographic key of arbitrary length. We will show that such a one-way permutation is a unitary operator that breaks its own symmetry, yielding a quantum cryptography 
protocol that is polynomially secure.

{\it{Preliminaries:}} Consider the Clauser-Horne-Shimony-Holt (CHSH) scenario \cite{Clauser1969}, where two spatially separated parties labeled Alice and Bob can 
accept binary inputs $a,x \in \{0,1\}$ and getting output bits $a$'$, x$'$ \in \{0,1\}$.
We can generate correlations between the output values and the input bits of a PR (Popescu-Rohrlich) box \cite{PopescuRohrlich1994} from a stochastic mechanism which 
depends on the temporal order of the inputs \cite{Popescu2006,Bub2012}. 
Suppose that the input $a$ is the temporal parameter, a control bit so that $ a$'$ $ occurs before $ x$'$ $. Then, for the group homomorphism 
$\{+1,-1,\times\} \mapsto \{0,1,\oplus\}$ so that its inverse is also a group homomorphism, 
the condition $a$'$:=0$ and $x$'$:=a\land x$ or $a$'$:=1$ and $x$'$=:1 \oplus a \land x$ produces the correlation
$a$'$\oplus x$'$:=a\land x$, where $\oplus$ is the addition modulo 2 ($XOR$) and 
the field's multiplication operation ($\times$) corresponds to the logical $AND$ function (Eq. \ref{box}). This mapping between two isomorphic groups can be written
as a 2-ary (total) function $g:(a,x) \mapsto [a$'$:=a,x$'$:= (f \in\{0,1\}) \oplus a \land x]$ defined for all possible input values, 
so that the communication system yields the PR correlation characterized by the following (conditional) probability distribution:

\begin{equation}
\label{box}
\mathcal Pr(\sfrac{a,x}{a',x'})=
\left \{
\begin{array}{cc}
\sfrac{1}{2},& a$'$\oplus x$'$:=a\land x  \\
0, & otherwise \\
\end{array}
\right.
\end{equation} whence, the input state of $g$ can only be guessed with negligible probability from its output state.

\subsection{Definition.} Let $ g:\{0,1\}^* \to \{0,1\}^*$ be a length-preserving 2-ary total function that is
easy to compute on every input but hard to invert given the image of a random input \cite{RabiSherman1997,HemaspaandraRothe1999}. The function $g$ is called strongly
one way if and only if the probability ${\mathcal Pr}$ of inverting $g$ is negligible (Eq. \ref{neg}). ${\mathcal Pr}$ is negligible if it
approaches zero faster than the multiplicative inverse of any positive polynomial:  
\begin{equation}
\label{neg}
{\mathcal Pr}_{\mathcal{A}(g) \in g^{-1}g}\in{\mathcal O}(\sfrac{1}{poly}),
\end{equation} where $\mathcal A$ is any probabilistic polynomial time algorithm \cite{Goldreich2004}.
In other words, a bad event that occurs with negligible probability ${\mathcal Pr}_{g^{-1}g\leftarrow g}<\sfrac{1}{poly}$
would be highly unlikely to occur even if we repeated the experiment polynomially many times. Otherwise, a function is called weakly one way if 
${\mathcal Pr}_{g^{-1}g\leftarrow g}>\sfrac{1}{poly}$, i.e., if an event that occurs with noticeable probability occurs almost always when the experiment
is repeated a polynomial number of times.
\subsubsection{Remark.} Inputs of $g(a,x)=(a,f(x)+ax)$ have $\leq 1$ siblings on average for any length-preserving $f$ and $a,x \in GF_{{2}^{\|x\|}}$
(see in \cite{Levin2003}):

i) The function $f$ is length preserving if for every $x \in \{0,1\}^*$ it holds that the
length of the input is the same as the length of the output. 

ii) The output $f(x)+ax$, where $a$ is a key bit, can be replaced by another hash function, a function that is used to map data of arbitrary sizes to data of fixed sizes.

iii) $\|x\|=length(x)$, and $GF_{2}$ is the Galois Field of two elements.

\subsubsection{Conjecture.} The above $g$ is one way, for any OWF (one-way function) $f$, and has the same (within a polynomial factor) security (see in \cite{Levin2003}):

i) This security scheme is provably secure if the probability of inverting $g$ grows asymptotically no faster than the multiplicative inverse of any positive polynomial
$p(x)$ for all large enough $\|x\|$.

ii) The polynomial $p(x)$ is positive over $GF_{{2}^{\|x\|}}$ if $p(x)>0$ for every $x \in GF_{{2}^{\|x\|}}$.
\section{Proof.}
The function $g(a,x)$ with $a,x \in GF_{{2}^{\|x\|}}$ is known as the universal one-way function. The question of whether one-way functions exist can be 
reduced to the question of whether this specific permutation is one way \cite{AroraBarak2009}.
\subsection{Definition.} Given a permutation of $n$ elements $g:\{1,...,n\}\to\{1,...,n\}$, its permutation matrix is a square binary (orthogonal) matrix which has exactly 
one entry of 1 in each row and each column and 0$'$s elsewhere. Its elements are $(n=\|x\|)$-bit arrays that can be represented as polynomials over the Galois fields 
$GF_{{2}^{\|x\|}}$ \cite{Bronshtein2004}. 
\subsubsection{Remark.}
For constructing a Galois extension of $GF_{2}$, e.g., the finite field $GF_{{2}^{length=3}}$ 
that represents the coordinates of the vertices defining a three-dimensional hypercube in which the sides are one unit in length, we need to choose an 
irreducible polynomial of degree 3 \cite{Stalling2011,Mullen2013}.

Let the Table \ref{tab1} below be the polynomial arithmetic modulo $x^3\oplus x \oplus 1$. Over the finite field with characteristic 2 (1+1=0), the field$'$s multiplication operation 
corresponds to the logical $AND$ gate, and the field$'$s addition operation corresponds to the logical $XOR$ gate. Hence, $g(a,x)=(a,f(x) \oplus (a \land x))$, and:

i) For $sibling=1$ (even input), $x=a$ implies that $f(x)=x [(I(x)]$; consequently, $g(a,x)=(a,x^{2} \oplus x)$ because $x=x^{2}$ over the finite field with characteristic 2 (see Table \ref{tab1}).

ii) For $sibling<1$ (odd input), $x \neq a$ implies that $f(x)=x^{2} \oplus 1 [NOT(x)]$; consequently, $g(a,x)=(a,x^{2} \oplus x \oplus 1)$ because $x \oplus 1=x^{2} \oplus 1$
over the finite field with characteristic 2 (see Table \ref{tab1}).

Note that $x^{2} \oplus x \oplus 1 > 0$ and $x^{2} \oplus x < 1$ for $x=\{0,1\}$. 

Thus, 
\begin{equation*}
g(0,0)=(0,0) \Rightarrow
\begin{bmatrix}
    g_{11}  & g_{12} & g_{13} &  g_{14} \\ 
    g_{21}  & g_{22} & g_{23} &  g_{24} \\
    g_{31}  & g_{32} & g_{33} &  g_{34} \\
    g_{41}  & g_{42} & g_{43} &  g_{44} 
\end{bmatrix}
\times
\begin{bmatrix}
1 \\ 
0 \\ 
0 \\
0
\end{bmatrix}
=
\begin{bmatrix}
    1 \\
    0 \\
    0 \\
    0
\end{bmatrix}
\therefore
\begin{bmatrix}
    g_{11} \\
    g_{21} \\
    g_{31} \\
    g_{41}
\end{bmatrix}
=
\begin{bmatrix}
    1 \\
    0 \\
    0 \\
    0
\end{bmatrix},
\end{equation*}
and
\begin{equation*}
g(1,0)=(1,1) \Rightarrow
\label{matrix_2}
\begin{bmatrix}
    1  & 0 & g_{13} &  g_{14} \\ 
    0  & 1 & g_{23} &  g_{24} \\
    0  & 0 & g_{33} &  g_{34} \\
    0  & 0 & g_{43} &  g_{44} 
\end{bmatrix}
\times
\begin{bmatrix}
0 \\ 
0 \\ 
1 \\
0
\end{bmatrix}
=
\begin{bmatrix}
    0 \\
    0 \\
    0 \\
    1
\end{bmatrix}
\therefore
\begin{bmatrix}
    g_{13} \\
    g_{23} \\
    g_{33} \\
    g_{43}
\end{bmatrix}
=
\begin{bmatrix}
    0 \\
    0 \\
    0 \\
    1
\end{bmatrix}.
\end{equation*}
as $g_{22}=1$ and $g_{12}=g_{32}=g_{42}=0$ for $g(0,1)=(0,1)$. In the same way, for $g(1,1)=(1,0)$, $g_{34}=1$ and $g_{14}=g_{24}=g_{34}=0$.

\begin{table}[h]
\centering
\caption{Logical operator precedence. For $a=NOT(x)$, $g(a,x)=(a,(x^{2} \oplus 1) \oplus ((x^{2}\oplus 1) \land x^{2}))$. Otherwise, $g(a,x)=(a,x \oplus (x \land x)).$}
\label{tab1}
\begin{adjustbox}{max width=\textwidth}
\begin{tabular}{
>{\columncolor[HTML]{EFEFEF}}c| 
>{\columncolor[HTML]{EFEFEF}}c 
>{\columncolor[HTML]{C0C0C0}}c 
>{\columncolor[HTML]{C0C0C0}}c 
>{\columncolor[HTML]{C0C0C0}}c 
>{\columncolor[HTML]{C0C0C0}}c 
>{\columncolor[HTML]{C0C0C0}}c 
>{\columncolor[HTML]{C0C0C0}}c 
>{\columncolor[HTML]{C0C0C0}}c 
>{\columncolor[HTML]{C0C0C0}}c}
      &                                   & \cellcolor[HTML]{EFEFEF}$000$ & \cellcolor[HTML]{EFEFEF}$001$ & \cellcolor[HTML]{EFEFEF}$010$     & \cellcolor[HTML]{EFEFEF}$011$ & \cellcolor[HTML]{EFEFEF}$100$ & \cellcolor[HTML]{EFEFEF}$101$  & \cellcolor[HTML]{EFEFEF}$110$  & \cellcolor[HTML]{EFEFEF}$111$    \\ \hline
      & (AND)                             & \cellcolor[HTML]{EFEFEF}$0$   & \cellcolor[HTML]{EFEFEF}$1$   & $x$                               & \cellcolor[HTML]{EFEFEF}$x+1$ & \color[HTML]{333333}$x^{2}$                          & \cellcolor[HTML]{EFEFEF}$x^{2}+1$ & \cellcolor[HTML]{EFEFEF}$x^{2}+x$ & \cellcolor[HTML]{EFEFEF}$x^{2}+x+1$ \\ 
$000$ & $0$                               & $0$                           & $0$                           & $0$                               & $0$                           & $0$                           & $0$                            & $0$                            & $0$                              \\
$001$ & $1$                               & $0$                           & $1$                           & $x$                               & $x+1$                         & $x^{2}$                            & $x^{2}+1$                         & $x^{2}+x$                         & $x^{2}+x+1$                         \\
$010$ & \cellcolor[HTML]{C0C0C0}$x$       & $0$                           & $x$                           & \cellcolor[HTML]{EFEFEF}$x^{2}$   & $x^{2}+x$                     & $x+1$                        & $1$                            & $x^{2}+x+1$                       & $x^{2}+1$                           \\
$011$ & $x+1$                             & $0$                           & $x+1$                         & $x^{2}+x$                         & $x^{2}+1$                     & $x^{2}+x+1$                      & $x^{2}$                           & $1$                            & $x$                              \\
$100$ & $x^{2}$                           & $0$                           & $x^{2}$                       & $x^{2}+1$                         & $x^{2}+x+1$                   & $x^{2}+x$                        & $x$                            & $x^{2}+1$                         & $1$                              \\
$101$ & \cellcolor[HTML]{C0C0C0}$x^{2}+1$ & $0$                           & $x^{2}+1$                     & $1$                               & $x^{2}$                       & \cellcolor[HTML]{EFEFEF}$x$   & $x^{2}+x+1$                       & $x+1$                          & $x^{2}+x$                           \\
$110$ & $x^{2}+x$                         & $0$                           & $x^{2}+x$                     & $x^{2}+x+1$                       & $1$                           & $x^{2}+1$                        & $x+1$                          & $x$                            & $x^{2}$                             \\
$111$ & $x^{2}+x+1$                       & $0$                           & $x^{2}+x+1$                   & $x^{2}+1$                         & $x$                           & $1$                           & $x^{2}+1$                         & $x^{2}$                    & $x+1$                           
\end{tabular}
\end{adjustbox}
\end{table}
\begin{table}[h]
\centering
\begin{adjustbox}{max width=\textwidth}
\begin{tabular}{
>{\columncolor[HTML]{EFEFEF}}c| 
>{\columncolor[HTML]{EFEFEF}}c 
>{\columncolor[HTML]{C0C0C0}}c 
>{\columncolor[HTML]{C0C0C0}}c 
>{\columncolor[HTML]{C0C0C0}}c 
>{\columncolor[HTML]{C0C0C0}}c 
>{\columncolor[HTML]{C0C0C0}}c 
>{\columncolor[HTML]{C0C0C0}}c 
>{\columncolor[HTML]{C0C0C0}}c 
>{\columncolor[HTML]{C0C0C0}}c }
      &                                 & \cellcolor[HTML]{EFEFEF}$000$ & \cellcolor[HTML]{EFEFEF}$001$ & \cellcolor[HTML]{EFEFEF}$010$     & \cellcolor[HTML]{EFEFEF}$011$ & \cellcolor[HTML]{EFEFEF}$100$   & \cellcolor[HTML]{EFEFEF}$101$       & \cellcolor[HTML]{EFEFEF}$110$     & \cellcolor[HTML]{EFEFEF}$111$       \\  \hline
      & (XOR)                           & \cellcolor[HTML]{EFEFEF}$0$   & \cellcolor[HTML]{EFEFEF}$1$   & $x$                               & \cellcolor[HTML]{EFEFEF}$x+1$ & \cellcolor[HTML]{EFEFEF}$x^{2}$ & $x^{2}+1$                           & \cellcolor[HTML]{EFEFEF}$x^{2}+x$ & \cellcolor[HTML]{EFEFEF}$x^{2}+x+1$ \\
$000$ & $0$                             & $0$                           & $1$                           & $x$                               & $x+1$                         & $x^{2}$                         & $x^{2}+1$                           & $x^{2}+x$                         & $x^{2}+x+1$                         \\
$001$ & $1$                             & $1$                           & $0$                           & $x+1$                             & $x$                           & $x^{2}+1$                       & $x^{2}$                             & $x^{2}+x+1$                       & $x^{2}+x$                           \\
$010$ & \cellcolor[HTML]{C0C0C0}$x$     & $x$                           & $x+1$                         & $0$                               & $1$                           & $x^{2}+x$                       & \cellcolor[HTML]{EFEFEF}$x^{2}+x+1$ & $x^{2}$                           & $x^{2}+1$                           \\
$011$ & $x+1$                           & $x+1$                         & $x$                           & $1$                               & $0$                           & $x^{2}+x+1$                     & $x^{2}+x$                           & $x^{2}+1$                         & $x^{2}$                             \\
$100$ & \cellcolor[HTML]{C0C0C0}$x^{2}$ & $x^{2}$                       & $x^{2}+1$                     & \cellcolor[HTML]{EFEFEF}$x^{2}+x$ & $x^{2}+x+1$                   & $0$                             & $1$                                 & $x$                               & $x+1$                               \\
$101$ & $x^{2}+1$                       & $x^{2}+1$                     & $x^{2}$                       & $x^{2}+x+1$                       & $x^{2}+x$                     & $1$                             & $0$                                 & $x+1$                             & $x$                                 \\
$110$ & $x^{2}+x$                       & $x^{2}+x$                     & $x^{2}+x+1$                   & $x^{2}$                           & $x^{2}+1$                     & $x$                             & $x+1$                               & $0$                               & $1$                                 \\
$111$ & $x^{2}+x+1$                     & $x^{2}+x+1$                   & $x^{2}+x$                     & $x^{2}+1$                         & $x^{2}$                       & $x+1$                         & $x$                                 & $1$                               & $0$                                
\end{tabular}
\end{adjustbox}
\end{table}
\vspace{1.0cm}
\hspace{-0.5cm}Therefore, the function $g$ is represented by the permutation matrix:

\begin{table}[ht]
\centering
\label{tab2}
\begin{tabular}{c|cccc}
                & $(a,x^{2} \oplus x)$ & $(a,x^{2} \oplus x \oplus 1)$ & $(a,x^{2} \oplus x)$ & $(a,x^{2} \oplus x \oplus 1)$ \\ \hline
$g_{(a<1,x=a)}$ & $1$                  & $0$                           & $0$                  & $0$                           \\
$g_{(a<1,x>a)}$ & $0$                  & $1$                           & $0$                  & $0$                           \\
$g_{(a>0,x<a)}$ & $0$                  & $0$                           & $0$                  & $1$                           \\
$g_{(a>0,x=a)}$ & $0$                  & $0$                           & $1$                  & $0$                          
\end{tabular}
\end{table}\hspace{-0.5cm}where the four columns correspond to the orthogonal basis $|00 \rangle$, $|01 \rangle$, $|11 \rangle$, $|10 \rangle$ of the Hilbert space ${\mathcal H}_{4}$. 

This covariance matrix form of standardized random variables denotes taht the average over the possible outcomes of all measurements may take on together 
according to the conditional joint probability distribution.
Such a matrix form of total 2-ary unitary operator $g$ is the controlled $NOT$ $(CNOT)$ function,
a two-qubit universal quantum gate defined for all possible input values, where $a$ is the control variable and $x$ is the target variable. 
\vspace{2cm}

Notice that for $a=x$, $CNOT|a \rangle \otimes |x \rangle \mapsto |a \rangle \otimes |x^{2} \oplus x \rangle$ 
and, for $a \neq x$, $CNOT|a \rangle \otimes |x \rangle \mapsto |a \rangle \otimes |x^{2} \oplus x \oplus 1 \rangle$, where 
the hashing $x^{2} \oplus x \oplus 1 = NOT(x^{2} \oplus x)$, with $x=\{0,1\}$.
The controlled $NOT$ gate acts on two qubits, and applies the $NOT$ gate 
$=
\begin{pmatrix}
0 & 1 \\
1 & 0
\end{pmatrix}$
to the target qubit $|x \rangle$ if the first (control) qubit, $|a \rangle $, 
is in state $|1 \rangle $. Otherwise, it applies the identity gate
$=
\begin{pmatrix}
1 & 0 \\
0 & 1
\end{pmatrix}$
if the the first qubit is in state $|0 \rangle $.

Considering the PR correlation given in Eq. \ref{box}, the 2-ary (total) function $g$ can be written as $CNOT(a,x)=(a$'$,x$'$)$, with
$ x$'$ =
\left \{
\begin{array}{cc}
\multicolumn{1}{l}{a\land x}, if $a=0$  \\
\multicolumn{1}{l}{1\oplus a\land x}, if $a=1$ \\
\end{array}
\right.$

Thus, $CNOT$ gate is completely specified by its truth table for $0\sim +1$ and $1\sim -1$:
\begin{table}[ht]
\centering
\label{tab2}
\begin{tabular}{c|c|c}
           &                     Input/Output                    &                   Input/Output                      \\ \hline
           &                                                     &                                                     \\ 
Target $=0$ & $|0\rangle|0\rangle \mapsto |+1\rangle|+1\rangle $  & $|1\rangle|0\rangle \mapsto |-1\rangle|-1\rangle $  \\
           &                                                     &                                                     \\ 
Target $=1$ & $|0\rangle|1\rangle \mapsto |+1\rangle|-1\rangle $  & $|1\rangle|1\rangle \mapsto |-1\rangle|+1\rangle $  \\ 
           &                                                     &                                                     \\ \hline
           &                     Control $=0$                    &                    Control $=1$                      \\
\end{tabular}
\end{table}

\hspace{-0.5cm}whence,
$NOT\begin{pmatrix}
+1  \\
-1 
\end{pmatrix}
=
\begin{pmatrix}
-1  \\
+1    
\end{pmatrix}$.
\subsection{Theorem.}
Let $(\Omega$, $\mathcal F$, $\mathcal Pr)$ be a Kolmogorov probability space with sample space $\Omega$, event space $\mathcal F$, and probability measure $ \mathcal Pr$.
Let $a,x$ be random variables; hence, the Clauser-Horne-Shimony-Holt (CHSH) inequality for correlations $|{\langle \cdot \rangle}_{(0,0)} + {\langle \cdot \rangle}_{(0,1)}
+ {\langle \cdot \rangle}_{(1,0)} - {\langle \cdot \rangle}_{(1,1)}| \leq 2$ holds in the Kolmogorov axiomatization \cite{Khrennikov2014}, where 
${\langle \cdot \rangle}_{(a,x \in GF_{2})}$ denotes the expectation values
for $x^{2}\oplus x$ and $x^{2}\oplus x \oplus 1$. 

Thus, $|{\langle x^{2} \oplus x \rangle}_{(0,0)}^{H} + {\langle x^{2} \oplus x \oplus 1 \rangle}_{(0,1)}^{H} + {\langle x^{2} \oplus x \oplus 1 \rangle}_{(1,0)}^{H}-{\langle x^{2} \oplus x \rangle}_{(1,1)}^{H}| \leq 2$,
measured on the Hadamard basis $H=\{|+\rangle,|-\rangle\}$ (see Eq. \ref{hadamard}). 
Therefore, the hidden (Markov) model, $|{\langle x^{2} \oplus x \oplus 1 \rangle}^{H}_{x=\{0,1\}}| \leq 1$, is the normalized upper bound 
to the correlation ${[x \oplus NOT(x)]}_{x=\{0,1\}}$ between two outcomes of the experiment (see Fig.1, below).
For this symmetric function, the variable $x=\{0,1\}$ (input state) is not directly visible, since $0$ and $1$ are equiprobable. However, the output dependent on the input state $x=\{0,1\}$ is visible.
According to reasoning assuming local hidden variable theory \cite{Sugiyama2014}, the correlation measure cannot exceed the value $2$, but there are four states of two qubits
which lead to the maximal value of $2\sqrt{2}$.

\tcbsidebyside[sidebyside adapt=left, blanker, sidebyside gap=0.38cm, 
               sidebyside align=top seam]{
\begin{tikzpicture}[scale=0.6]
\draw [thick,-latex, white] (0,0)--(4.3,0) node [right, above]{};
\draw [thick,-latex] (0,0)--(0,8.3) node [above]{\small{$NOT(x)$}};
\coordinate (X2) at (3,0);
\coordinate (P1) at (0,1);
\coordinate (P3) at (0,7);
\coordinate (P2) at (0,4);
\coordinate (b) at (3,1);
\coordinate (a) at (3,7);
\coordinate (c) at (4.3,4);
\coordinate (d) at (3.4,4);
\node at (X2) [below] {};
\node at (P2) [left] {\small{$0$}};
\node at (P1) [left] {\small{$-|1 \rangle $}};
\node at (P3) [left] {\small{$+|1 \rangle $}};
\draw [dashed] (a) node [above right] {$$}--(b);
\draw [dashed] (P1)--(b) node [below, right] {\small{$\frac{1}{\sqrt{2}}|0\rangle - \frac{1}{\sqrt{2}}|1\rangle $}};
\draw [dashed] (P3)--(a) node [above, right] {\small{$\frac{1}{\sqrt{2}}|0\rangle + \frac{1}{\sqrt{2}}|1\rangle $}};
\draw [-latex] (P2)--(b) node [above,sloped,midway] {\scriptsize{$x \oplus NOT(x)$}};
\draw [dashed] (P2)--(d) node [above] {\small{$|0\rangle$}};
\draw [dashed] (P2)--(d) node [below] {\small{$|0\rangle$}};
\draw [thick,-latex] (P2)--(c) node [right] {\small{$x$}};
\draw [-latex] (P2)--(a) node [above,sloped,midway] {\scriptsize{$x \oplus NOT(x)$}};
\node at (b) [above right]{$$};
\tkzDrawArc[rotate,color=black](P2,b)(180)
\end{tikzpicture}
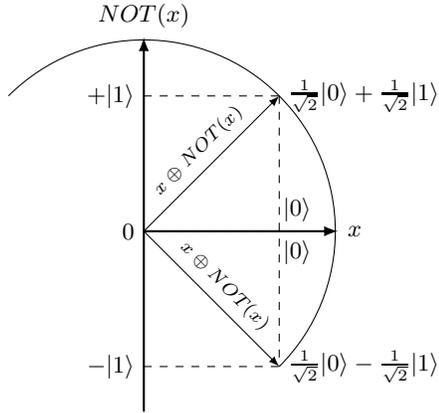
}{\small{Fig.1. The absolute value of the symmetric difference $-1 \leq {\langle x \oplus NOT(x)\rangle}_{x=\{0,1\}}^{H} \leq +1$ is the normalized Euclidean metric from 
the point in two-dimensional rectangular space 
$\bigg[|0 \rangle = 
\begin{pmatrix}
1 \\
0
\end{pmatrix}\bigg]$ 
$\oplus$ 
$\bigg[|1 \rangle =
\begin{pmatrix}
0 \\
1
\end{pmatrix}\bigg]$
$=$ 
$\frac{1}{\sqrt{2}}
\begin{pmatrix}
1 \\
1
\end{pmatrix}$ to the origin of the Cartesian coordinate system of the complex plane. In the Argand diagram, one can see that 
${[x \oplus NOT(x)]}_{x=\{0,1\}}$ and its complex conjugate has the same absolute value. Hence, $|{\langle{x^{2} \oplus x \oplus 1} \rangle}^{H}_{x=\{0,1\}}| \leq 1$, where $NOT$ logic gate can
be simulated by addition (modulo 2) operation, $ NOT(x) = x^{2} \oplus 1 $, with $x \in GF_{{2}^{||x||}}$. Namely, one quantum bit can contain at most one classical bit of 
information, which is in accordance with the Holevo bound \cite{Holevo1973}}.} 

\subsubsection{Remark.}
Let the controlled $NOT$ function be $CNOT(|a \rangle \otimes |x \rangle) \mapsto |a \rangle \otimes |a \oplus x \rangle$, with the output values $\{+1,-1\}$
and its inputs, $\{0,1\}$, so that $0 \sim +1$ and $1 \sim -1$. The correlations ${[x^{2} \oplus x]}_{x=\{0,1\}}$ and ${[x^{2} \oplus x \oplus 1]}_{x=\{0,1\}}$
are used to realize the Bell states, and their conjugates, $\Phi^{\pm}$ and $\Psi^{\pm}$. 

Consider the Hadamard basis $\{|+\rangle,|-\rangle\}$ of a one-qubit register given by the
size-2 (discrete Fourier transform) DFT 
$\begin{pmatrix}
\multicolumn{1}{c}{1} & \multicolumn{1}{r}{1} \\
\multicolumn{1}{c}{1} & \multicolumn{1}{c}{-1}
\end{pmatrix}{|x\rangle}_{x=\{0,1\}}$:
\begin{equation}
\label{hadamard}
{|x \rangle}_{x=0,1} \xrightarrow{H} \frac{1}{\sqrt{2}}[(-1)^{x}|x \rangle + |1-x \rangle].
\end{equation} 

The following quantum circuits, $QXOR(x^{H},I(x))$ and $QXOR(x^{H},NOT(x))$, evolve the four inputs, $|a \rangle|x \rangle$ for $a,x=\{0,1\}$, into the four entangled states of two qubits:
\vspace{2cm}

i) For $|x \rangle $ = $|a \rangle $, and $x^{2}=x$ over $GF_{{2}^{||x||}}$, we have:
\begin{figure}[ht]
  \centerline{
    \begin{tikzpicture}[thick]
    \tikzstyle{operator} = [draw,fill=white,minimum size=1.5em] 
    \tikzstyle{phase} = [fill,shape=circle,minimum size=5pt,inner sep=0pt]
    \tikzstyle{empty} = [draw,fill=white,shape=circle,minimum size=5pt,inner sep=0pt]
    \tikzstyle{surround} = [fill=white!10,thick,draw=black,rounded corners=2mm]
    \node at (-0.27,0) (q1) {$\ket{x=0}$};
    \node at (-0.48,-1) (q2) {$\ket{x^2}$};
    \node[operator] (op11) at (1,0) {H} edge [-] (q1);
    \node[empty] (op21) at (2,-1) {$+$} edge [-] (q2);
    \node[phase] (phase11) at (2,0) {} edge [-] (op11);
    \node[empty] (phase12) at (2,-1) {$+$} edge [-] (op21);
    \draw[-] (phase11) -- (phase12);
    \node[phase] (phase21) at (2,0) {} edge [-] (phase11);
    \node[empty] (op24) at (2,-1) {$+$} edge [-] (phase12);
    \node (end1) at (3,0) {} edge [-] (phase21);
    \node (end2) at (3,-1) {} edge [-] (op24);
    \draw[decorate,decoration={brace},thick] (3,0.2) to
	node[midway,right] (bracket) {\hspace{0.2cm}$ QXOR[\frac{1}{\sqrt{2}}(|0 \rangle + |1 \rangle),|0 \rangle] = \frac{1}{\sqrt{2}}|0\rangle |0\rangle + \frac{1}{\sqrt{2}}|1\rangle |1\rangle = |\Phi^{+} \rangle $.}
        (3,-1.2);
    \end{tikzpicture}
}
  \centerline{
    \begin{tikzpicture}[thick]
    \tikzstyle{operator} = [draw,fill=white,minimum size=1.5em]
    \tikzstyle{phase} = [fill,shape=circle,minimum size=5pt,inner sep=0pt]
    \tikzstyle{empty} = [draw,fill=white,shape=circle,minimum size=5pt,inner sep=0pt]
    \tikzstyle{surround} = [fill=white!10,thick,draw=black,rounded corners=2mm]
    \node at (-0.2,0) (q1) {$\ket{x=1}$};
    \node at (-0.398,-1) (q2) {$\ket{x^2}$};
    \node[operator] (op11) at (1,0) {H} edge [-] (q1);
    \node[empty] (op21) at (2,-1) {$+$} edge [-] (q2);
    \node[phase] (phase11) at (2,0) {} edge [-] (op11);
    \node[empty] (phase12) at (2,-1) {$+$} edge [-] (op21);
    \draw[-] (phase11) -- (phase12);
    \node[phase] (phase21) at (2,0) {} edge [-] (phase11);
    \node[empty] (op24) at (2,-1) {$+$} edge [-] (phase12);
    \node (end1) at (3,0) {} edge [-] (phase21);
    \node (end2) at (3,-1) {} edge [-] (op24);
    \draw[decorate,decoration={brace},thick] (3,0.2) to
        node[midway,right] (bracket) {\hspace{0.2cm}$ QXOR[\frac{1}{\sqrt{2}}(|0 \rangle - |1 \rangle),|1 \rangle] = \frac{1}{\sqrt{2}}|0\rangle |1\rangle - \frac{1}{\sqrt{2}}|1\rangle |0\rangle = |\Psi^{-} \rangle $.}
        (3,-1.2);
    \end{tikzpicture}
  }
\end{figure}

ii) For $|x \rangle \neq |a \rangle $, and $x^{2} \oplus 1 = x \oplus 1 $ over $GF_{{2}^{||x||}}$, we have:
\vspace{0.3cm}
\begin{figure}[ht]
  \centerline{
    \begin{tikzpicture}[thick]
    \tikzstyle{operator} = [draw,fill=white,minimum size=1.5em]
    \tikzstyle{phase} = [fill,shape=circle,minimum size=5pt,inner sep=0pt]
    \tikzstyle{empty} = [draw,fill=white,shape=circle,minimum size=5pt,inner sep=0pt]
    \tikzstyle{surround} = [fill=white!10,thick,draw=black,rounded corners=2mm]
    \node at (-0.2,0) (q1) {$\ket{x=0}$};
    \node at (-0.13,-1) (q2) {$\ket{x^2 \oplus 1}$};
    \node[operator] (op11) at (1,0) {H} edge [-] (q1);
    \node[empty] (op21) at (2,-1) {$+$} edge [-] (q2);
    \node[phase] (phase11) at (2,0) {} edge [-] (op11);
    \node[empty] (phase12) at (2,-1) {$+$} edge [-] (op21);
    \draw[-] (phase11) -- (phase12);
    \node[phase] (phase21) at (2,0) {} edge [-] (phase11);
    \node[empty] (op24) at (2,-1) {$+$} edge [-] (phase12);
    \node (end1) at (3,0) {} edge [-] (phase21);
    \node (end2) at (3,-1) {} edge [-] (op24);
    \draw[decorate,decoration={brace},thick] (3,0.2) to
        node[midway,right] (bracket) {\hspace{0.2cm}$ QXOR[\frac{1}{\sqrt{2}}(|0 \rangle + |1 \rangle),|1 \rangle] = \frac{1}{\sqrt{2}}|0\rangle |1\rangle +\frac{1}{\sqrt{2}}|1\rangle |0\rangle =|\Psi^{+} \rangle $.}
        (3,-1.2);
    \end{tikzpicture}
  }
  \centerline{
    \begin{tikzpicture}[thick]
    \tikzstyle{operator} = [draw,fill=white,minimum size=1.5em]
    \tikzstyle{phase} = [fill,shape=circle,minimum size=5pt,inner sep=0pt]
    \tikzstyle{empty} = [draw,fill=white,shape=circle,minimum size=5pt,inner sep=0pt]
    \tikzstyle{surround} = [fill=white!10,thick,draw=black,rounded corners=2mm]
    \node at (-0.21,0) (q1) {$\ket{x=1}$};
    \node at (-0.13,-1) (q2) {$\ket{x^2 \oplus 1}$};
    \node[operator] (op11) at (1,0) {H} edge [-] (q1);
    \node[empty] (op21) at (2,-1) {$+$} edge [-] (q2);
    \node[phase] (phase11) at (2,0) {} edge [-] (op11);
    \node[empty] (phase12) at (2,-1) {$+$} edge [-] (op21);
    \draw[-] (phase11) -- (phase12);
    \node[phase] (phase21) at (2,0) {} edge [-] (phase11);
    \node[empty] (op24) at (2,-1) {$+$} edge [-] (phase12);
    \node (end1) at (3,0) {} edge [-] (phase21);
    \node (end2) at (3,-1) {} edge [-] (op24);
    \draw[decorate,decoration={brace},thick] (3,0.2) to
        node[midway,right] (bracket) {\hspace{0.2cm}$ QXOR[\frac{1}{\sqrt{2}}(|0 \rangle - |1 \rangle),|0 \rangle] = \frac{1}{\sqrt{2}}|0\rangle |0\rangle - \frac{1}{\sqrt{2}}|1\rangle |1\rangle =|\Phi^{-} \rangle $.}
        (3,-1.2);
    \end{tikzpicture}
  }
\end{figure}

This simple network applies the Hadamard gate given by Eq. \ref{hadamard} to the first wire and XORs
the randomized first wire into the second wire yielding the maximally entangled states $|\Phi^{\pm}\rangle$ = $(|0\rangle\pm |1 \rangle) QXOR |0 \rangle = 
QXOR(|0\rangle \pm |1\rangle,|0\rangle)=QXOR(|0\rangle,|0\rangle) \pm QXOR(|1\rangle,|0\rangle)$ 
and $|\Psi^{\pm}\rangle$ = $(|0\rangle \pm |1 \rangle) QXOR |1 \rangle = QXOR(|0\rangle \pm |1\rangle,|1\rangle)=QXOR(|0\rangle,|1\rangle) \pm QXOR(|1\rangle,|1\rangle)$,
where the normalization constant is omitted.
The quantum exclusive $OR$ operation ($QXOR$) corresponds to $CNOT$ gate that
flips the second (target) qubit if the first (control) qubit is $|1 \rangle $ and does nothing if the control qubit is
$|0 \rangle $. 

\vspace{6.0cm}
Let $\Psi^{+}=||\Psi^{+}\rangle|$ and $\Phi^{-}=||\Phi^{-}\rangle|$ be the expectations of the correlations ${[x \oplus NOT(x)]}_{x=\{0,1\}}$, after the quantum circuit to 
perform a Hadamard transform followed by controlled $NOT$ gate on the input values.
From the sum of $\Psi^{+}$ and $\Phi^{-}$, we can write down the set of four correlations in the experiment:
\begin{table}[h]
\centering
\begin{tabular}{lll}
$\hspace{0.15cm}{[x \oplus NOT(x)]}_{x=0}\hspace{0.462cm}        \xrightarrow{H-CNOT} $ & $\frac{1}{\sqrt{2}}||01\rangle + |10\rangle|$                           & \multirow{2}{*}{(+)} \\
                                                                                     &                                                                           &                       \\ 
$\hspace{0.15cm}{[x \oplus NOT(x)]}_{x=1}\hspace{0.462cm}        \xrightarrow{H-CNOT} $ & $\frac{1}{\sqrt{2}}||00\rangle - |11\rangle|$                           &                      \\ \hline
                                                                                     &                                                                           &                       \\
$2{[x \oplus NOT(x)]}_{x=\{0,1\}} \xrightarrow{H-CNOT} $ & $\frac{1}{\sqrt{2}}||01\rangle + |10\rangle| + \frac{1}{\sqrt{2}} ||00\rangle - |11\rangle|$ &                          (=)                  
\end{tabular}
\end{table}

\hspace{-0.5cm}whence, ${\langle x \oplus NOT(x) \rangle}_{x=\{0,1\}}^{H-CNOT} \ge \frac{1}{2\sqrt{2}}||00\rangle + |10\rangle| + |10\rangle - |11\rangle|$,
once by subadditivity property (triangle inequality), $||00\rangle + |10\rangle| + |10\rangle - |11\rangle| \le ||01\rangle + |10\rangle| + ||00\rangle - |11\rangle|$.
As $x \oplus NOT(x)=1$ for $x \in GF_{{2}^{||x||}}$, we have that the sum of correlations is
$S \le 2\sqrt{2}$, where $ S=||00 \rangle + |01 \rangle+ |10 \rangle - |11 \rangle|$ on the Hadamard basis $H=\{|+\rangle,|-\rangle\}$. 
Thus, the mathematical formalism shows that quantum correlations go up to Tsirelson's bound of the CHSH inequality.

Notice that the exclusive disjunction $x \oplus NOT(x)= x^{2} \oplus x \oplus 1$ is the polynomial representation of the power set ${\mathcal F}^{\Omega}$ of the universal set $\Omega = \{x$'$,x$''$,x$'''$\}$.
Its subsets are $\{\}:=0$, $\{x$'$\}:=x^{2}$, $\{x$''$\}:=x$, $\{x$'''$\}:=1$, $\{x$'$,x$''$\}:=x^{2}\oplus x$,
$\{x$'$,x$'''$\}:=x^{2}\oplus 1$, $\{x$''$,x$'''$\}:=x \oplus 1$ and $\{x$'$,x$''$,x$'''$\}:=x^{2}\oplus x \oplus 1$,
namely the Cartesian coordinates of the Euclidean space $\mathbb{R}^{3}$.
\vspace{0.5cm}

The set of the subsets of the ${\mathcal F}^{\Omega}$ ordered by inclusion composes a poset -- a partially ordered set in which binary relations as
$\le$ hold for some pairs of elements of the set, but not for all --, where the irreducible polynomial $x^{2}\oplus x \oplus 1$ over $GF_{{2}^{||x||}}$ dependents on itself for its existence.
In this ontological dependency defined on the
three-dimensional space model of the physical universe\footnote[1]{Verify that $x^{x} \oplus x \oplus 1$ is
the universal set (of everything) in the computational knowledge engine https://www.wolframalpha.com/input/?i=(x+and+x)+xor+x+xor+1.
Verify also that $ NOT(x^{2} \oplus x \oplus 1) = x^{2} \oplus x$ is the empty set $\{\emptyset \}$ in https://www.wolframalpha.com/input/?i=(x+and+x)+xor+x. 
Another interesting point about the polynomial $x^{2} \oplus x \oplus 1$ can be seen in \cite{deCastro2016}.}, the basis elements of a bigger Hibert 
space ${\mathcal{H}}_{0}$, which is a superset of the conventional
Hibert space $\mathcal{H}$, evolve into basis elements \cite{Hooft2009} in accordance with 
the Hasse diagram shown in Fig.2. Consequently, asymptotic behaviour is associated with ${\mathcal F}^{\Omega}$, once the universal set is large enough.
\vspace{0.2cm}
\tcbsidebyside[sidebyside adapt=left, blanker, sidebyside gap=0.3cm, 
               sidebyside align=top seam]{
\begin{psmatrix}[mnode=circle,colsep=0.6,rowsep=0.6]
& [name=8] 111 \\[-0.1\psyunit]
[name=5] 110 & [name=6] 101 & [name=7] 011\\
[name=2] 100 & [name=3] 010 & [name=4] 001\\[-0.1\psyunit]
& [name=1] 000
\ncline{8}{5}\ncline{8}{6}\ncline{8}{7}
\ncline{5}{3}\ncline{5}{2}
\ncline{6}{2}\ncline{6}{4}
\ncline{7}{3}\ncline{7}{4}
\ncline{1}{2}\ncline{1}{3}\ncline{1}{4}
\ncline[offset=5pt, linewidth=0.5\pslinewidth, arrows=->, nodesep=5pt]{1}{2}
\ncline[offset=5pt, linewidth=0.5\pslinewidth, arrows=->, nodesep=5pt]{1}{3}
\ncline[offset=5pt, linewidth=0.5\pslinewidth, arrows=->, nodesep=5pt]{1}{4}
\ncline[offset=5pt, linewidth=0.5\pslinewidth, arrows=->, nodesep=5pt]{2}{5}
\ncline[offset=5pt, linewidth=0.5\pslinewidth, arrows=->, nodesep=5pt]{2}{6}
\ncline[offset=5pt, linewidth=0.5\pslinewidth, arrows=->, nodesep=5pt]{3}{5}
\ncline[offset=5pt, linewidth=0.5\pslinewidth, arrows=->, nodesep=5pt]{3}{7}
\ncline[offset=5pt, linewidth=0.5\pslinewidth, arrows=->, nodesep=5pt]{4}{6}
\ncline[offset=5pt, linewidth=0.5\pslinewidth, arrows=->, nodesep=5pt]{4}{7}
\ncline[offset=5pt, linewidth=0.5\pslinewidth, arrows=->, nodesep=5pt]{5}{8}
\ncline[offset=5pt, linewidth=0.5\pslinewidth, arrows=->, nodesep=5pt]{6}{8}
\ncline[offset=5pt, linewidth=0.5\pslinewidth, arrows=->, nodesep=5pt]{7}{8}
\end{psmatrix}
}{\small{Fig.2. Ontology chart \cite{Kecheng2000,Stamper2004}(Hasse diagram) of the partially ordered set of all subsets of $\{x$'$,x$''$,x$'''$\}$. 
The  subsets $\{\}=(000)$, $\{x$'$\}=(100)$, $\{x$''$\}$, $\{x$'''$\}=(001)$, $\{x$'$,x$''$\}=(110)$, 
$\{x$'$,x$'''$\}=(101)$, $\{x$''$,x$'''$\}=(011)$ and $\{x$'$,x$''$,x$'''$\}=(111)$ are basis elements that evolve into $\{x$'$,x$''$,x$'''$\}$,
and represent the coordinates of the vertices defining a 3-D hypercube.
By the Cantor's first uncountability proof, such an Euclidean space has the same cardinality of the unit interval $[0,1]$. 
The segment $[0,1]$ is a subset of ${\mathbb{R}}$, and it has the cardinality of the continuum. 
Therefore, the edge of the 3-D hypercube whose side has length one unit is equal to the cube root of its volume  
$\sqrt[3]{1}=
\left \{
\begin{array}{cc}
\multicolumn{1}{l}{1}  \\
-\sfrac{1}{2} \pm i\sfrac{\sqrt{3}}{2} \\
\end{array}
\right.$,
where the Galois conjugates $-\sfrac{1}{2} \pm i\sfrac{\sqrt{3}}{2}$ are zeros of the minimal polynomial $poly(x)$.
Every minimal polynomial is irreducible over $GF_{{2}^{||x||}}$.}} 

\subsubsection{Remark.}
Measuring the first bit of the pairs $|{\Psi}^{+} \rangle $ and $|\Phi^{-} \rangle $ in the computational basis yields a $0$ or $1$ with probability $\sfrac{1}{2}$. 
Likewise, measuring its second bit yields the same outcome with the same probability. Therefore, measuring one bit of the maximally entangled two-qubit Bell states  
yields a random outcome. Hence, we can rewrite the EPR pairs $|{\langle x^{2} \oplus x \oplus 1 \rangle}_{x=\{0,1\}}^{H}| \leq 1$ as a Markov's inequality 
$ poly(x){\mathcal Pr}_{poly(x)=0,1} \leq {\langle x^{2} \oplus x \oplus 1 \rangle}_{GF_{{2}^{||x||}}}$, where the sample space $poly(x)=|x^{2} \oplus x \oplus 1|$ is 
the indicator random variable $1_{\mathcal F}: \Omega \mapsto \mathbb{R}$ defined by $1_{\mathcal F}(x)=1$ if $x \in {\mathcal F}^{\Omega} $, otherwise, $1_{\mathcal F}(x)=0$. 
The measure ${\mathcal Pr}_{poly(x)=0,1}$ is the probability of factoring, ${\mathcal Pr}_{poly(x)=0}$, or non-factoring, ${\mathcal Pr}_{poly(x)=1}$,
the Bell states $|{\Psi}^{+} \rangle $ and $|\Phi^{-} \rangle $ generated by $QXOR(x^{H},NOT(x))$. Recall that the polynomial $poly(x)$ is factorable over $GF_{{2}^{||x||}}$ if 
$poly(x)=u(x)v(x)$ with both non-constant polynomials $u(x)$ and $v(x)$ $\in GF_{{2}^{||x||}}$, otherwise, $poly(x)$
is irreducible. If the degree of $poly(x)$ is $2$, then $poly(x)$ is a non-factorable polynomial over the finite field $GF_{{2}^{||x||=3}}$ if and only if $p(x)$ has no root in $GF_{{2}^{||x||=3}}$,
i.e., $poly(x)=1$ for $x=\{0,1\}$. 

For the sake of simplicity, from now on, we will replace $\oplus \mapsto (+)$, and will use $\oplus $ only where strictly necessary to ensure the coherence of the operation. 
The notation will also be simplified, once the computational basis and Hadamard basis are isomorphic.

\subsection{Theorem.}
The probability of factoring $poly(x)$, ${\mathcal Pr}_{poly(x)=0}$, is negligible if and only if the product $poly(x){\mathcal Pr}_{poly(x)=0}$ 
approaches $0$ asymptotically for any positive polynomial $poly(x)>0$. (See a proof of this theorem for negligible functions in \cite{Goldreich2004}).

\subsubsection{Remark.}
Any positive polynomial over $GF_{{2}^{||x||}}$ is reduced to the irreducible polynomial $poly(x)=x^{2} \oplus x \oplus 1 > 0$. Thus, 
$poly(x)$ is almost surely non-factorable, since the probability of non-reducing it, ${\mathcal Pr}_{poly(x)=1}$, is equal to one.

Let the factorization of $poly(x)$ be a tail event $E \in \mathcal{F}$ in the probability space $(\Omega, \mathcal F, \mathcal Pr)$ that happens almost surely 
if ${\mathcal Pr}[E]=1$. Equivalently, $E$ occurs almost surely if the probability of $E$ not occurring is ${\mathcal Pr}[E^{c}]=0$, where $E^{c}$ is the complementary 
event (Kolmogorov$'$s zero--one law, see proof in \cite{JacodProtter2004}).

Consequently, $poly(x){\mathcal Pr}_{poly(x)=0} < 1$, because the probability of factoring $poly(x)$ vanishes for $x=\{0,1\}$. As a result, ${\mathcal Pr}_{poly(x)=0}$ is negligible, 
once it approaches $0$ quickly as ${\mathcal Pr}_{poly(x)= 0}<\frac{1}{poly(x)}$, where $poly(x)=x^{2}+x+1 > 0$ and the field's addition operation $(+)$ corresponds to the
exclusive $OR$ logical operation $(\oplus)$ given the random input $x=\{0,1\}$. 
Notice that we can map the elements of the Hadamard basis to the computational basis using the group homomorphism $\{+1,-1,\times\} \mapsto \{0,1,\oplus\}$
so that its inverse is also a group homomorphism. Then, the
exclusive disjunction $ x \oplus NOT(x) = x^{2} \oplus x \oplus 1 $ can be rewritten as $ x^{2} \oplus x \oplus 1 := l$'$ \land \neg l$''$ $, once the field's multiplication operation
corresponds to the logical AND operation over the field of two elements. It is not difficult to see that for $ l$'$ = l$''$ = l$'''$ $, 
$ l$'$ \land \neg l$'$ = (l$'$ \lor l$''$ \lor l$'''$)\land(\neg l$'$ \lor \neg l$''$ \lor \neg l$'''$)$ can be written as 3CNF (conjunctive normal form) clauses, $(l$'$ \lor l$''$ \lor l$'''$)\land (l$'$ \lor l$''$ \lor \neg l$'''$)\land(l$'$ \lor \neg l$''$ \lor l$'''$) \land (l$'$ \lor \neg l$''$ \lor \neg l$'''$) \land (\neg l$'$ \lor l$''$ \lor l$'''$) \land (\neg l$'$ \lor l$''$ \lor \neg l$'''$) \land (\neg l$'$ \lor \neg l$''$ \lor l$'''$) \land (\neg l$'$ \lor \neg l$''$ \lor \neg l$'''$)$, 
which is unsatisfiable. As a result, factoring the polynomial $poly(x)$ over $GF_{{2}^{||x||=3}}$ is as hard as solving the Boolean satisfiability problem (SAT):
the variables of the Boolean formula above can be consistently replaced by the values TRUE or FALSE in such a way that the formula evaluates to TRUE?

Try this Fortran code to see:

PROGRAM RANDOM

\hspace {1cm} LOGICAL $x$,$y$,$z$ 

\hspace {1cm} $y$ = .NOT. $x$

\hspace {1cm} $z$ = .TRUE.

\hspace {1cm} $x$ .neqv. $y$ = $z$

WRITE(*,*) $x$

END

Is there another programming language able to solve this problem? 

There is no deterministic way even if we repeat the experiment polynomially many times, since ${\mathcal Pr}_{poly(x)=0}$ is negligible over the Boolean ring of all subsets of $x$.

\subsubsection{Remark.}
The question above can directly be replaced by the problem of whether $poly(x)>0$ with any reasonable probability distribution on its inputs can be factored in polynomial time on average.
\cite{Cook2017,Levin1986,Impagliazzo1995}. 
\vspace{0.5cm}

Time complexity analysis: Let the bigger Hilbert space ${\mathcal{H}}_{0} \supseteq \mathcal{H}$ be the same size as the set of all subsets. ${\mathcal{H}}_{0}$ has the cardinality of the continuum; therefore,
the (discrete) distribution $\sfrac{1}{poly(x)}$ over ${GF_{2}}$, where every element\footnote[2] {The unitary operator $NOT^{2}(x)=I(x)$: 
$\begin{bmatrix}
    0  & 1 \\
    1  & 0
\end{bmatrix}
\times
\begin{bmatrix}
    0 & 1 \\
    1 & 0
\end{bmatrix}
=
\begin{bmatrix}
    1 & 0 \\
    0 & 1
\end{bmatrix}$. Thus $(x^{2}+1)^{2}=x^{2}$, hence, $x^{2}+1=\pm x$, and we have $poly(x)=x^{2}+x+1$ or $poly(x=-x)=x^{2}-x+1$, once every element $x=x^{2}$ of $GF_{2}$ satisfies the property 
$x \oplus x=0$.} $x=-x$, converges to the bell-shaped (continuous) curve of the probability density function of the Cauchy distribution (its left tail is shown in Fig.3), 
with integral principal value (P.V.) equal to $\sfrac{1}{2}$
and probability ${\mathcal Pr}_ {x \in {\mathcal F}^{\Omega}}$ given by  
\begin{equation}
\label{cauchy}
\frac{1}{N}\int_ {- \infty}^{+\infty} {\frac{1}{poly(x)}dx}=1,
\end{equation}
where the normalizing constant $N=2\pi\sqrt{\sfrac{1}{3}}$.
\tcbsidebyside[sidebyside adapt=left, blanker, sidebyside gap=0.3cm,
               sidebyside align=top seam]{
    \begin{tikzpicture}[
    declare function={gamma(\z)=2.506628274631*sqrt(1/\z)+ 0.20888568*(1/\z)^(1.5)+ 0.00870357*(1/\z)^(2.5)- (174.2106599*(1/\z)^(3.5))/25920- (715.6423511*(1/\z)^(4.5))/1244160)*exp((-ln(1/\z)-1)*\z;},
    declare function={gammapdf(\x,\k,\theta) = 1/(\theta^\k)*1/(gamma(\k))*\x^(\k-1)*exp(-\x/\theta);},
    declare function={cauchypdf(\x,\mu,\gamma) = 1/(pi*\gamma*(1+((\x-\mu)/\gamma)^2));} 
]
\begin{axis}[
  legend pos=north west,
  legend style={draw=none},
  no markers, domain=0:10, samples=100,
  axis lines*=center, xlabel=$x$, ylabel=$PDF's$,
  every axis y label/.style={at=(current axis.above origin),anchor=south},
  every axis x label/.style={at=(current axis.right of origin),anchor=west},
  height=6cm, width=7.7cm,
  xtick={33.3}, ytick=\empty,
  xticklabels={$\sfrac{1}{2}$}
  =false, clip=false, axis on top,
  grid = major
  ]
\legend{\tiny{$x \sim$ Gaussian},\tiny{$x \sim$ Cauchy}}
    \addplot [dashed, smooth, domain=0:33] {gammapdf(x,3.5,13)};
    \addplot [thick, smooth, domain=0:50] {cauchypdf(x,33.5,17.05)};
\coordinate (a) at (320,4);
  \node (b) at (250,50) {\tiny{Cauchy P.V.}};
  \draw (a) edge[out=20, in=-95,->] (b);
   \end{axis}
   \end{tikzpicture}
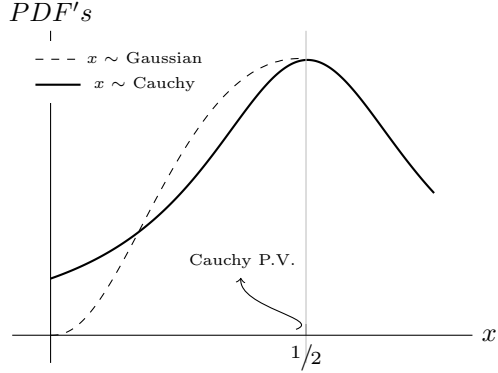
}{\small{Fig.3. The probability density function of the Cauchy distribution (Eq. \ref{cauchy}) can be written as
\begin{equation*}
PDF_{Cauchy}={\frac{1}{\pi}}{\frac{\alpha}{(x-\mu)^{2}+{\alpha}^{2}}},
\end{equation*} where $\alpha$ is the half width at half maximum and $\mu$ is
the statistical median. $PDF_{Cauchy}$ is similar in appearance to Gaussian curve, however,
the values for away from P.V. are much more likely than they would be with a Normal distribution since its tails drop off much more slowly.}}

As the sample space is large enough and the input $x$ is a Cauchy-distributed random variable,
the polynomial $x^{2}+ x+1>0$, with $x  \in GF_{{2}^{||x||}}$, is asymptotically almost surely
a hard core, once the presence of the heavy extreme values in the Cauchy distribution means that the average value does not converge to a fixed value. 
The Cauchy distribution is a heavy-tailed distribution belonging to the subexponential class whose probability density function decreases at a polynomial rate as 
$x \to -\infty $ and $x \to +\infty $, as opposed to an exponential rate. (The polynomial $x^{2}+ x + 1 = x+(x^{2}+1) \in GF_{{2}^{||x||}}$, where input $x$ in $x^{2}+1$ is also   
a (standard) Cauchy-distributed random variable (Witch of Agnesi). 

Consequently, the probability of factoring the predicate $poly(x)=x^{2}+ x+1>0$ for $x=\{0,1\}$ --- which is identical to 
finding a way that 3CNFSAT evaluates to TRUE --- 
is subexponentially bounded making the factorization of $poly(x)$ an NP-complete problem, which is in accordance with the exponential time hypothesis \cite{ImpagliazzoPaturi1999}.
Thus, the running time $T(x)$ of any cryptanalysis algorithm to factorize $poly(x)$ on inputs of size $||x||$ grows faster than polynomial time, since 
3CNFSAT cannot be decided in the subexponential class.

Considering that every exponential time algorithm takes longer than a subexponential time algorithm as $||x||$ increases, then,
the running time of any algorithm to factorize $poly(x)$ is order of complexity $T(x)=2^{\mathcal{O}(x)}$ in big $\mathcal{O}$-notation.

It is straightforward to see that the expectation of the squared deviation (variance) of the random variable $x$ can be radically altered by the extremes of the Cauchy distribution.
Hence, if the variance is unpredictable, the maximally entangled state $x^{2}+ x+1>0 \in GF_{{2}^{||x||}}$ 
is asymptotically almost surely a hard-core predicate which is easy to compute given $x$, but is hard to compute $x$ given its output of a single bit. This predicate
(hidden Markov model) provides every one-way functions with a hidden bit of the same security. It yields a "perfect" random generator (PRG)
with maximum entropy probability \cite{Hastad1999} from any one-way bijection, since the input $x$ computed from the output can only be guessed with probability $\sfrac{1}{2}$.
This maximum min-entropy --- the smallest entropy measure in the family of R\'{e}nyi$'$s entropies --- is a measure of how correlated the state $poly(x)$ is. 

\subsection{Theorem.}
Let one-to-one correspondence $g$ be a function defined as $g(a,x)=(a,h(x))$, where the length of $a$ is the same as that of $x$, and $h(x)=f(x)+ax$ over $GF_{{2}^{||x||}}$. 
The Boolean inner product ${\left\langle a\oplus x\right\rangle}_{a \neq x}$ provides a one-way function with a hidden bit of the same security. 
(See the proof of this theorem in \cite{GoldreichLevin1989,Goldreich2004,Hastad1999,Levin1993}.

\subsubsection{Remark.}
The hard-core predicate of $g$ is the parity function of a random subset of the inputs of $g$. 
If $g$ has a hard-core predicate $h(x)$, then it must be strongly one way. Hence, the probability of inverting $g$, ${\mathcal Pr}_{g^{-1}g \gets g}$,
is the same probability of factoring the hard-core $h(x)$.
Then, the probability of inverting $g$ is negligible because the probability of factoring the maximally
entangled state $h(x)=f(x)+ax = poly(x) \in GF_{{2}^{||x||}}$, with $x \neq a$, is less than $\frac{1}{x^{2}+x+1>0}$. Consequently, ${\mathcal Pr}_{g^{-1}g \gets g}$ 
approaches zero faster than $\frac{1}{poly(x)>0}$ given the random input $x=\{0,1\}$, where $poly(x) = x^{2} \oplus x \oplus 1$ is the only positive 
polynomial\footnote[3]{Evidently, any positive polynomial over $GF_{{2}^{||x||}}$ is reduced to $x^{2} \oplus x \oplus 1$.} among the
$2^{3}=8$ polynomials over $GF_{{2}^{||x||=3}}$.  

\subsection{Theorem.}
If $P \neq NP$\footnote[4]{The $P \stackrel {?}{=} NP$ problem is to determine whether every language accepted by some nondeterministic algorithm in
polynomial time is also accepted by some (deterministic) algorithm in polynomial time \cite{Cook1971}. $P \neq NP$  if and only if a total 2-ary one-way functions exists
\cite{RabiSherman1997,HemaspaandraRothe1999}.}, then, some strongly non-invertible functions are invertible (see proof in \cite{Hemaspaandra2006}).
\vspace{0.5cm}
\subsubsection{Corollary.}
Let $g$ be the controlled NOT gate, and its unitary (and Hermitian) matrix written in the form:

\begin{equation}
U_{CNOT}=
\begin{bmatrix}
\label{matrix_1}
    1 & 0 & 0 &  0 \\
    0 & 1 & 0 &  0 \\
    0 & 0 & 0 &  1 \\
    0 & 0 & 1 &  0
\end{bmatrix},
 \hspace{0,2cm}U_{CNOT}^{2}=
\begin{bmatrix}
    1 & 0 & 0 &  0 \\
    0 & 1 & 0 &  0 \\
    0 & 0 & 1 &  0 \\
    0 & 0 & 0 &  1
\end{bmatrix}.
\end{equation}

The liner operator $U_{CNOT}=U_{CNOT}^{-1}=U_{CNOT}^{T}$ is orthogonal. Hence, $g$ is involutory: a bijective map that is its own inverse, i.e., a mirror symmetry 
because when it is applied twice in succession, every state returns to its original value. A bijective function from a set to itself is a permutation \cite{RichterGebert2011}. 

\subsubsection{Remark.}
It is straightforward to see that strongly non-invertible functions are invertible from the definition itself of one-way functions. 
(See a thermodynamic approach of one wayness \cite{deCastro2014} in input-saving machines \cite{Bennett1989,Ozawa2002}).

Consider $g$ defined on pairs of strings of the same length, so that $g(a,x)=(a,f(x)\oplus x)$ (pg. 94 in \cite{Goldreich2004}).

Thus, it is self-evident that the functions
\begin{equation}
g=
\begin{bmatrix}
\label{matrix_1}
    $f(x)= I(x)$ &      0_{2}        \\
          0_{2}    & $f(x)=NOT(x)$ 
\end{bmatrix}
\end{equation}
and $f(x)$ have information-theoretic security within the same polynomial factor.

Let the hard-core $h(x)$ in $g$ be a permutation $f$'$(x)=f(x)\oplus x$, where $f$ is any (length-preserving) one-way function. 
As the output of the $XOR$ bitwise operation $f$'$(x)=f(x)\oplus x$ is true if and only if the inputs are not alike; otherwise, the output is false, $f(x)$ in Eq.6 can only be represented by the 
polynomials $f(x)=x$ $[I(x)]$ or $f(x)=x\oplus1$ $[NOT(x)]$ over $GF_{{2}^{||x||}}$.

Let $GF_{{2}^{||x||}}$ a field and $f(x)$ a polynomial in $GF_{{2}^{||x||}}$. If deg[$f(x)]=1$, then, $f(x)$ is non-factorable over $GF_{{2}^{||x||}}$.
This is obvious because the polynomial $f(x)$ is factorable over $GF_{{2}^{||x||}}$ if and only if $f(x)=u(x)v(x)$ with both non-constant polynomials $u(x)$ and $v(x)$.
If $f(x)=u(x)v(x)$ for some $u(x),v(x) \in GF_{{2}^{||x||}}$, then, deg[$f(x)$]$=$deg[$u(x)$]+deg$[v(x)]$. However, deg[$u(x)$]+deg[$v(x)$] are nonnegative integers over the
integral domain $GF_{{2}^{||x||}}$, hence, one of the degrees must be $0$. Thus, either $u(x)$ or $v(x)$ must be a constant polynomial.
It follows that $f(x)$ is almost surely non-factorable\footnote[5]{The functions $I(x) = x \equiv x^{2}$ and $NOT(x) = x \oplus 1 \equiv x^{2} \oplus 1$ hold. However, $x$ and $x \oplus 1$
are irreducible (non-factorable), while $x^{2}$ and $x^{2} \oplus 1$ are reducible (factorable) over $GF_{{2}^{||x||}}$.
The probability density function $\sfrac{1}{(x^{2}+1)} $ is the Witch of Agnesi, a heavy-tailed distribution
belonging to the subexponential class (see time complexity analysis in Remark 2.3.1), while the reciprocal random variable $\sfrac{1}{x}$ is an exponential random
variable (exponential of the uniform random variable $x=\{0,1\}$).} over $GF_{{2}^{||x||}}$. Hence, $f(x)=x$ or $f(x)=x\oplus1$ are 
length-preserving one-way functions over $GF_{{2}^{||x||}}$, and any length-preserving one-way function over $GF_{{2}^{||x||}}$ is reduced to them. 
As a result, $f$'$(x)=x^{2}\oplus x$ or $f$'$(x)=x^{2}\oplus x\oplus1$, where $x=x^{2}$ over $GF_{{2}^{||x||}}$. 

Notice that the polynomial $x^{2}\oplus x$ is factorable over $GF_{{2}^{||x||}}$ because it outputs $0$ for $x=\{0,1\}$. Otherwise, 
the polynomial $x^{2}\oplus x \oplus 1$ is almost surely non-factorable over $GF_{{2}^{||x||}}$ because the probability of factoring it is negligible. 
Consequently, $f$'$(x)$ is weakly one way for every even input and strongly one way for every input odd (see Remark 2.2.1).
However, the polynomial $x^{2}\oplus x \oplus 1 = (x^{2}\oplus x)\oplus 1$ outputs $1$ for $x=\{0,1\}$, then, by symmetry, $x^{2}\oplus x= 1\oplus 1$ yielding $x^{2}\oplus x=0$ for $x=\{0,1\}$. 
As a result, the exclusive disjunctions $x^{2}\oplus x$ and $x^{2}\oplus x\oplus 1$ are deducible from each other, since $XOR$ operation is involutory. Therefore,
every strongly one-way function is also weakly one way, once any positive polynomial over
$GF_{{2}^{||x||}}$ and any polynomial zero over $GF_{{2}^{||x||}}$ is reduced to $x^{2}\oplus x \oplus 1=1$ and 
$x^{2}\oplus x=0$ over $GF_{{2}^{||x||=3}}$, respectively. 
As there is a one-to-one correspondence between a complex number and its complex conjugate, the equipollence between
the polynomials $x^{2}\oplus x$ and $x^{2}\oplus x \oplus 1$ is self-evident, since Bell states and its conjugates
$|{\phi}^{+}\rangle/|{\phi}^{-}\rangle$ and $|{\psi}^{-}\rangle/|{\phi}^{+}\rangle$ are generated by $x^{2}\oplus x$ and $x^{2}\oplus x \oplus 1$, respectively (see detail in Remark 2.2.1).

Recall that the three-dimensional space $\{x$'$,x$''$,x$'''$\}$ is represented by the Hasse diagram shown in Fig.2. In that ontological chart, 
a state is partially ordered with another state, where in every such pair of states 
we will label the first as $Alice$ and the second as $Bob$. There are $2^{3}$ possible combinations of such states given in the Table \ref{tab2} below:

\begin{table}[ht]
\caption{Polynomial representation of the pairs $Alice$ and $Bob$.}
\label{tab2}
\centering
\begin{tabular}{cc|cc|c}
   Alice   &                                 &   Bob    &                                     & Probability          \\ 
   x$'$x$''$x$'''$    &        Polynomial               &   x$'$x$''$x$'''$    &        Polynomial              &                           \\ \hline 
   $111$   &\multicolumn{1}{l}{$x^{2}+x+1$}  &  $000$   &\multicolumn{1}{l}{$0$}         & ${\mathcal Pr}^{1}$       \\ 
   $110$   &\multicolumn{1}{l}{$x^{2}+x$}    &  $001$   &\multicolumn{1}{l}{$1$}         & ${\mathcal Pr}^{2}$       \\
   $101$   &\multicolumn{1}{l}{$x^{2}+1$}    &  $010$   &\multicolumn{1}{l}{$x$}         & ${\mathcal Pr}^{3}$       \\
   $100$   &\multicolumn{1}{l}{$x^{2}$}      &  $011$   &\multicolumn{1}{l}{$x+1$}       & ${\mathcal Pr}^{4}$       \\
   $011$   &\multicolumn{1}{l}{$x+1$}        &  $100$   &\multicolumn{1}{l}{$x^{2}$}     & ${\mathcal Pr}^{5}$       \\
   $010$   &\multicolumn{1}{l}{$x$}          &  $101$   &\multicolumn{1}{l}{$x^{2}+1$}   & ${\mathcal Pr}^{6}$       \\
   $001$   &\multicolumn{1}{l}{$1$}          &  $110$   &\multicolumn{1}{l}{$x^{2}+x$}   & ${\mathcal Pr}^{7}$       \\
   $000$   &\multicolumn{1}{l}{$0$}          &  $111$   &\multicolumn{1}{l}{$x^{2}+x+1$} & ${\mathcal Pr}^{8}$ 
\end{tabular}
\end{table}

\hspace{-0,5cm}where ${\mathcal Pr}^{i}$, with $i=1,...,8$, is the probability of a of a specific combination occurring in the sample space including all possible combinations. 
The bit arrays, $Alice$ and $Bob$, are polynomials $p_{i}(x) \in GF_{{2}^{||x||=3}}=\{0,1\}$ (as shown in Table \ref{tab1}, Remark 2.1.1).

Taking into account the Sakurai's Bell inequality \cite{Sakurai1994}, we can have that
${\mathcal Pr}^{3} + {\mathcal Pr}^{4} \le {\mathcal Pr}^{3} + {\mathcal Pr}^{4} + {\mathcal Pr}^{2} + {\mathcal Pr}^{7}$ holds, 
where the probabilities are always nonnegative ${\mathcal Pr}^{i} = |p_{i}(x)|$, 
with every polynomial $p_{i}(x) = \frac{ {\langle p_{i}(x) \rangle}}{|x^{2} + x + 1|}$. The polynomial $x^{2} + x + 1$ is the powerset of all possible combinations over $GF_{{2}^{||x||=3}}$

Therefore, the modulo 2 arithmetic is (i) $|(x^{2} + 1) + x^{2}| \le |(x^{2} + 1) + x^{2} + (x^{2} + 1) + 1|$ for $Alice$, 
and (ii) $|x + (x + 1)| \le |x + (x + 1) + 1 + (x^{2} + x)|$ for her logical complement, $Bob$. By subadditivity, we have
$|x^{2} + x + 1| \le |x^{2} + x|$ for both configurations, where $x=x^{2}$ over $GF_{{2}^{||x||}}$.

As the polynomials $x^{2} + x + 1$ and $x^{2} + x$ are logically deducible from each other over the finite field with characteristic 2, then, the inequality is reversed. Namely, 
$|x^{2} + x| \le |x^{2} + x + 1|$ because $x^{2} + x$ is ground set of $x^{2} + x + 1$ in the partially ordered set $\{x$'$,x$''$,x$'''$\}$.

Consider, now, the Cantor-Schr\"oder-Bernstein theorem below:

Theorem: Given two sets $\mathcal A$ ($Alice$) and $\mathcal B$ ($Bob$). If $ t$'$ $: $\mathcal A \to \mathcal B $ and $ t$''$ $: $\mathcal B \to \mathcal A $ are both injections, then, 
there exists a bijective function $\mathcal A \sim \mathcal B$ (see proof in \cite{HalmosGivant2008,Hinkis2011}).
\vspace{0.3cm}

Thus, $x^{2} + x$ can be exchanged by $x^{2} + x + 1$ so that $|x^{2}+ x +1| \le |x^{2}+ x +1|$, since
there is a one-to-one correspondence between the polynomials for $x=\{0,1\}$ (they are equivalent). In fact, the powerset $x^{2} + x + 1$ over $GF_{{2}^{||x||=3}}$ has 
cardinality strictly less than or equal to itself cardinality, as shown in Fig.2.
Consequently, the multiplicative inverse $\frac{1}{poly(x)} \ge \frac{1}{poly(x)}$ holds, where $poly(x)=|x^{2}+ x +1|$.

As $]0,1[ \subseteq \mathbb{R}$ and $[0,1] \subseteq \mathbb{R}$ have the same cardinality, the multiplicative inverse $\frac{1}{poly(x)} < \frac{1}{poly(x)}$ for $x=\{0,1\}$, obviously.
This condition implies that the strongly one-way function $x^{2} + x + 1 \in GF_{{2}^{||x||}}$ --- polynomial 
whose (negligible) probability of factoring it approaches zero quickly --- exists because the weakly one-way function
$x^{2} + x \in GF_{{2}^{||x||}}$ exists --- polynomial whose (noticeable) probability of factoring it does not approach zero too quickly. The reverse is also true, 
since every strongly one-way function is also weakly one-way \cite{Hemaspaandra2006,Zimand2004}. 
Therefore, $x^{2}+x$ is separable (classically correlated) because the probability of factoring it is not less than $\frac{1}{poly(x)}$.
Otherwise, $x^{2}+x+1$ is entangled (or non-separable) because the probability of factoring it is not greater than $\frac{1}{poly(x)}$.

This multiplicative inverse polynomial distance between an entangled state and the separable set reduces the separability criterion 
in bidirectional quantum controlled schemes \cite{Bayer2006,Li2016,LiLi2016} to an NP-hard problem \cite{Gharibian2010}.

In accordance with Fig.1, the size-2 (discrete Fourier transform) DFT over the finite field with characteristic 2 generates the unit vector $|x + NOT(x) \rangle$
with coordinates $(\frac{1}{\sqrt{2}},\frac{1}{\sqrt{2}})$ making a ${45}^{\circ}$ angle with the 
axes in the plane. Hence, the probability amplitude (wave function) is equal to the reciprocal of $\sqrt{2}$ computed over $\mathbb{R}$.
This number satisfies $\sin({45}^{\circ})$, therefore, $[\sin({45}^{\circ})]^{2}=[\frac{1}{poly(x)}]^{2}$, where $poly(x)=|x^{2}+x+1|$. 
Consequently, the trigonometric inequality $\frac{1}{2}[\frac{1}{poly(x)}]^{2} \le \frac{1}{2}[\sin(22.5^{\circ})]^{2} + \frac{1}{2}[\sin(22.5^{\circ})]^{2}$ holds. As a result, 
$0.2500 \le 0.1464$, and the inequality is maximally violated for the values predicted for the ``Bell test angles \cite{Bell2004}.''
However, $\frac{1}{poly(x)} \ge \frac{1}{poly(x)}$ over $GF_{{2}^{||x||}}$, and considering that the ring of integers modulo 2 consists only of idempotent elements, we have that
$[\frac{1}{poly(x)}]^{2}=\frac{1}{poly(x)}$, hence, $\frac{1}{2}[\frac{1}{poly(x)}]^{2} \ge \frac{1}{2}[\sin(22.5^{\circ})]^{2} + \frac{1}{2}[\sin(22.5^{\circ})]^{2}$. As a result, 
$0.2500 \ge 0.1464$, and the inequality is not violated for the values predicted for the ``Bell test angles''. This logical loophole \cite{Hess2004,Raedt2012,Hess2015,Raedt2016,Hooft2016,Khrennikov2016} 
stems directly from the existence of one-way functions, since the weak one-way function, $x^{2}+x$, can be used to produce the strong one-way function, $x^{2}+x+1$ in 
accordance with the amplifying hardness (Yao’s $XOR$ Lemma) \cite{AroraBarak2009,YAO1977}.

\subsection{Conclusion.}
Levin and Goldreich \cite{GoldreichLevin1989,Levin1993} proved that the hard core of the universal one-way function $g$ is a hidden bit (deterministic) model
able to generate randomness. (See also pseudorandom generator theorems \cite{Goldreich2004,Hastad1999}). Here, our one-way protocol showed that the 
(pseudo)randomness -- necessary and sufficient condition -- to buid the secure scheme $g$ is achieved, since the Bell inequality can be reduced to polynomial inequality 
$|x^{2}+x+1| \le |x^{2}+x+1|$. Whence, the asymptotic security $|x^{2} + x + 1|{\mathcal Pr}_{g^{-1}g} \le 1$ for ${\mathcal Pr}_{g^{-1}g} \le 1$ is obtained from a 
deterministic process over the Boolean ring of all subsets of $x$. Conversely, there is no deterministic process that produces $|x^{2}+ x + 1|{\mathcal Pr}_{g^{-1}g} > 1 $ 
for ${\mathcal Pr}_{g^{-1}g} \le 1$, although both conditions are deducible from each other. Such an ``equalness-of-strength'' shows that the problem of determining whether a 
given state is entangled or separable is at least as hard as the hardest problems in $NP$.

\section*{acknowledgements}
The author wishes to express thanks to his colleagues at Embrapa, E. H. dos Santos and J. G. Minto Neto who have discussed and collaborated for a long time during work. 
The author would also like to thank the anonymous reviewers for their valuable comments and suggestions.

\end{document}